\documentclass[%
reprint,
amsmath,amssymb,
aps,
]{revtex4-2}

\usepackage{braket}
\usepackage{graphicx}
\usepackage{dcolumn}
\usepackage{bm}
\usepackage{xcolor}

\begin{document}
\preprint{APS/123-QED}
	
\title{An Adaptive Genetic Algorithm for determining optimal structures for atomic clusters}	
\author{Brandon Willnecker}
\author{Mervlyn Moodley}
\affiliation{School of Chemistry and Physics, University of KwaZulu-Natal, Westville Campus, Private Bag X54001, Durban, 4000, South Africa}
	
\date{\today}
	
\begin{abstract}
The implementation of adaptive genetic algorithms (AGA) for optimization problems has proven to be superior than many other methods due to its nature of producing more robust and high quality solutions. Considering the complexity involved in many-body simulations, a novel AGA is proposed for applications to such systems and is specifically used to determine the lowest energy structures of various sized atomic clusters. For demonstrative purposes, we apply our method to various sized Lennard-Jones clusters and show that our results are more accurate than those found in the literature employing different methods.
\end{abstract}
	
\maketitle
\section{Introduction}
\label{sec1} 
The simulation of many-body systems are infamous for being algorithmically complex and computationally expensive. The many-body problem is a general name for a collection of physical problems concerned with the properties of microscopic systems composed of many interacting particles \cite{feynman} \cite{fetter}. The many interactions between the particles of the system create quantum correlations or entanglements which greatly complicate the composite wave-function for the system and as a consequence makes the problem highly computationally expensive. The computational cost to solve a classical $N$-body problem grows as $O(N^2)$ and for a quantum $N$-body problem it grows as $O(2^N)$. A typical macroscopic system contains on the order of $N=10^{23}$ particles and so this makes $N$-body physics one of the most computationally intensive fields in science. Problems of this type occur in many areas such as Bose-Einstein condensates, superfluids, quantum chemistry, atomic physics, molecular physics, nuclear physics, quantum chromodynamics and many others. The problem is therefore prevalent in both physics and chemistry and so it is interesting, if not necessary, to find new techniques or adapt existing ones to more efficiently solve these problems.

The use of Monte Carlo methods, both the classical and quantum versions, have proved the mainstay of solving these type of problems for many years \cite{monte_carlo_1,monte_carlo_2}. Recently, there has been much interest in the use of artificial neural networks (ANN) \cite{artificial_nn} and other machine learning methods \cite{machine_learning_in_phys}. Here the many-body wave function,$\psi(\vec{r_1},..\vec{r_N},w_1,..,w_N)$, is encoded by a set of parameters in a neural network, where $(w_1,...,w_N)$ is the parameter set. These parameters are then optimized or "trained" using a reinforcement learning procedure \cite{reinforcement_learning}. During this procedure, the parameters of the neural network are adjusted in such a way as to minimize a "cost function". This function gives a numerical way of judging how close to optimal a given set of parameters is. This technique was used in Ref.\cite{artificial_nn} to study the Ising and Heisenberg models in both one and two dimensions and achieved results at the same level as the best-known variational ansatz \cite{results_for_ising}. The high accuracy obtained for the unitary dynamics of quantum many-body systems suggests that neural-network–based approaches can be successfully used to solve the quantum many-body problem, not only for ground-state properties but also for the time evolution of the systems.

Simulated annealing (SA) is a meta-heuristic method used to approximate the global extreme of a specified function \cite{simulated_annelaing}. Simulated annealing in many-body physics involves constructing an ansatz wave function of the system in terms of a set of parameters. This is similar to ANN, however, the optimization is done by making random steps through the parameter space. The step is then accepted or rejected using a Boltzmann weight, $e^{-\Delta E/T}$ \cite{simulated_annealing_bw}. $\Delta E$ is the change in the system energy calculated from the current set of parameters and the new set. $T$ is a temperature-like parameter that controls the bias in accepting a given step. A low temperature will lead to steps being taken in a direction that will reduce the energy and hence move towards the goal of finding the ground state wave function. This temperature parameter changes over time according to an "annealing schedule" \cite{simulated_annealing_temp_schedule}. The simulation is started with a high temperature to allow a larger portion of the parameter space to be explored. This value is lowered each iteration to narrow down the search. The function describing the change in $T$ is problem dependent and can be changed easily. At the end of the simulation the current point in the parameter space is taken to be the best approximation to the global minimum. The convergence of this algorithm is discussed in Ref.  \cite{simulated_annealing_convergence}. These parameters can then be used to find the wave function and the ground state energy of the system. An implementation of this algorithm is presented in Ref. \cite{simulated_annealing_example} to solve the one and two dimensional short range Ising spin glass problem.

Basin-hopping (BH) or Monte-Carlo minimization (MCM) are so far the most reliable algorithms in chemical physics to search for the lowest-energy structure of an atomic cluster and macromolecular systems \cite{Masao_Iwamatsu_basin_hopping,basin_hopping}. It follows the procedure of repeatedly making random steps in the search space, performing local optimizations and then accepting or rejecting the new state based on a minimized function value. This algorithm is very useful in high-dimensional spaces such as the configuration space of a many particle system such as molecules. This algorithm was utilized by Wales and Doye\cite{basin_hopping} to determine the lowest energy structures for Lennard-Jones clusters with up to 110 particles. As done by many other researchers similarly employing the Lennard-Jones potential, the results from our work are compared to the latter results for particle numbers in the range $2 \leq N \leq 20$.

A Genetic algorithm (GA) \cite{GA_toolbox} is a meta-heuristic inspired by the process of natural selection. They are designed to solve various problems such as the travelling salesman problem and the knapsack problem which are too costly to solve using deterministic algorithms. GAs operate repeatedly on a population of potential solutions labelled as individuals, which are encoded in strings called chromosomes, by applying the principle of "survival of the fittest" to obtain increasingly accurate approximations to the solution \cite{mitchell,review_of_selection_methods}. At each step or generation, a new set of possible solutions is generated by selecting individuals based on fitness and "breeding" them together. This process is continued until a maximum number of generations is reached or a solution within a certain tolerance level is found. GAs are most commonly used to produce highly accurate solutions to a wide variety of optimization and search problems. Some of these problems include scheduling \cite{GA_scheduling_problem}, chess problems \cite{GA_chess_problem} and production optimization problems from economics \cite{GA_economics_prolem}.

Genetic algorithms can be used as a valuable tool to study the lowest energy states of physical systems. The only requirements are that the system needs to be represented in a meaningful way, either by a set of particle locations or a parametrized wave function, such that the GA operators can be applied and a fitness function can be defined. In most cases this can be the energy of the system or an appropriate function of the energy. The GA could provide a number of advantages over other algorithms. These include the fact that a GA works on a large set of possible solutions instead of a single possible solution. This leads to a greater chance of migrating towards the optimal solution with each iteration. The genetic algorithm proposed in this paper is designed to find the global minimum of various sized particle clusters interacting via the Lennard-Jones potential. It should be pointed out that various authors \cite{Deaven, Romero, Barron, Muller} have used some form of GA, different to what is discussed in this paper, to investigate specifically Lennard-Jones clusters. It is not the purpose of this paper to discuss the similarities or differences of these works but in retrospect to say that our method is more accurate since its is only limited by machine precision.

In Sec.~\ref{sec2}, we describe in some detail the structure and implementation of the genetic algorithm for an $N$-particle system. The results produced by the GA are presented in Sec.~\ref{sec3} and compared to the results found in Ref. \cite{basin_hopping}. Several difficulties are also discussed as well as how this GA can be improved upon and adapted to work for general potentials. Finally, a conclusion follows in Sec.~\ref{sec4}.

\section{Structure of the GA for an $N$-particle system}
\label{sec2}
The constituents of the genetic algorithm for an $N$-particle system include the population, the fitness function that is used to evaluate how good the candidate solution is at solving the problem, selection as well as the two operators, crossover and mutation. We also analyse our implementation of the GA and the various features designed to improve the rate of convergence for this particular problem. 
\subsection{Population and Initialisation}
The individuals of the GA are the cluster structures and so the chromosomes can be represented by $x\in\mathbb{R}^{3N}$ where $\mathbb{R}^{3N}$ specifies the configuration space of an $N$-particle system in $\mathbb{R}^3$.

Since the energy of a cluster is invariant under translation, rotation and permutation of the particle cluster, the configuration space $\mathbb{R}^{3N}$ contains degenerate state vectors that could negatively affect the performance of the GA since two seemingly different cluster structures are actually the same structure in terms of identical particle arrangement. The translation invariance of the clusters can be removed by imposing that the centre of mass (COM) remains fixed at the origin at all time. The rotation and permutation invariances are more difficult to remove.

The shape of the potential can be used to our advantage when generating the initial population. There is a minimum distance where the potential is positive at smaller distances. If the initial particle clusters can be generated whereby they are at least this specified distance apart then the resulting energy will be at most zero. The Poisson disk dampling algorithm in three dimensions \cite{poisson_sampling}, with some modifications, is used to accomplish this. The steps of the algorithm are as follows:
\begin{itemize}
\item Step 1: Generate a 3-dimensional grid to store the positions of the particles. This grid speeds up particle look up. The size of the grid cells is set to $\frac{r_0}{\sqrt{3}}$, where $r_0$ is the minimum distance between particles that we wish to enforce. Choosing this to be the cell size allows only one particle per cell.
\item Step 2: Randomly generate a position for the first particle. This is inserted into the grid and into a list. This list is used to store "active particles" that can be used to determine the position of the next particle.
\item Step 3: While the particle number is less than $N$, choose a particle from the active list, say the $i^{th}$ particle, and generate a new particle in the region between $r_0$ and $2r_0$ around the particle, check the neighbourhood to see if any intersections (inter-particle distance less than $r_0$) occurs and if there are no intersections then add this particle to the active list. If intersections are found then remove the $i^{th}$ particle from the active list.
\end{itemize}

\subsection{Fitness Function}
The fitness of a given individual is determined by a fitness function which gives a value that describes how suitable the individual is at being a solution to the problem \cite{intro_to_GA}. In our case, the energy of the given cluster is used as the fitness. However, since the target energy is negative and one would like lower energies to correspond to higher fitness values, a normalised fitness function is defined,
\begin{equation}
F(x_i) = \frac{E_{max}-E(x_i)}{E_{max}-E_{min}}
\end{equation}
where $E(x_i)$ is the total energy of the $i^{th}$ individual in the generation. $E_{max}$ and $E_{min}$ are the largest and smallest energy values from the generation respectively. Since the fitness function is normalised, the best performing individual will have a fitness of 1 and the worst performing individual will have a fitness of 0. The individuals can then be selected for crossover (discussed below) using their fitness value.

This AGA will work for any pair-wise potential function and we have chosen the Lennard-Jones potential for demonstrative purposes. For this potential, the total energy of an individual (particle cluster) is given by,
\begin{equation}
E = \sum_{i=1}^{N-1}\sum_{j=i+1}^{N} V(r_{ij})
\end{equation}
where $V(r_{ij})$ is the Lennard-Jones potential given by,
\begin{equation}
V(r_{ij}) = \frac{1}{r_{ij}^{12}} -2\frac{1}{r_{ij}^6}
\end{equation}
where $r_{ij}$ is the inter-particle distance between the $i^{\rm th}$ and $j^{\rm th}$ particle in the cluster. The potential is written using reduced units\cite{merv}.

\subsection{Selection}
The fitness level of an individual determines the probability of being selected for mating. A higher fitness level means a higher chance of being selected and so a higher chance of passing its genes on to the next generation \cite{review_of_selection_methods}. The Boltzmann selection method is implemented in our AGA. This method produces a distribution of clusters that corresponds to canonical distributions found in statistical physics and so results in a more optimized search. Given the fitness of an individual, $F(x_i)$, the probability to be selected for crossover is given by,
\begin{equation}
p(x_i) = \frac{1}{Z}e^{-\beta F(x_i)}
\end{equation}
where
\begin{equation}
Z = \sum_{j=1}^{N}e^{-\beta F(x_i)}
\end{equation}
and $\beta$ is a free parameter. $Z$ is the normalisation constant for the probabilities and $N$ is the number of individuals in the population. In this form, $\beta$ appears to be inverse temperature and $Z$ appears to be the partition function in statistical physics, but this is not the case. Only a very small subset of systems is considered and so this normalisation constant, $Z$, can not play the role of a partition function. However, this inverse temperature parameter can play a similar role to its thermodynamic counterpart because increasing $\beta$ (decreasing $T$) will increase the likelihood of lower energy systems being chosen which is analogous to a system being more likely in a lower energy state at low temperatures. This fact is used to change $\beta$ accordingly to guide the AGA into producing higher fitness (lower energy) individuals.

Elitism \cite{intro_to_GA} is used in this AGA to ensure that the fitter individuals are not lost during the selection stage. This is where a proportion of the fittest individuals is moved to the next generation with no modifications. This proportion amount is called the elitism factor. An elitism factor of $0.1$ (10$\%$ of the population) was used for the implementation of this algorithm.

\subsection{Crossover}
Genetic operators \cite{genetic_operators} manipulate the genes of the chromosomes under the assumption that the gene code from certain individuals will produce fitter individuals over a number of generations. The crossover operator \cite{crossover} will take two individuals called parents and produce two children by combining the genes from both chromosomes. In other words, we implement a function that takes in two parents and produces a child that will be placed in the next generation. This function needs to combine the chromosomes of the parents such that the child gets the "better qualities" of both parents on average. This ensures that the next generation will be fitter than the previous.

In terms of the Lennard-Jones cluster problem, we need to form a set of particles from the union of parent A and B, and then create a child cluster from these particles. A graph theoretical approach can be used to reformulate this problem. 

Consider the following graph made by combining the particles from two $N=3$ clusters as shown in Figure 1.
\begin{figure}[h]
	\centering
	\includegraphics[scale=0.65]{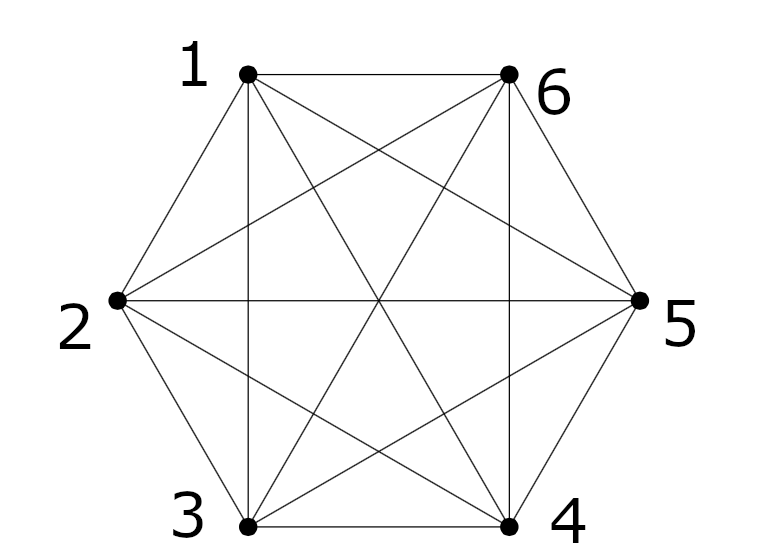}
	\caption{A complete graph with 6 vertices and 15 edges formed from the union of the two parent clusters.}
	\label{fig:completegraph}
\end{figure}
Each vertex represents a particle from the union and each edge represents the bond between the particles. The creation of a child cluster now entails finding an $N$-vertex sub graph. This can be done in three steps:
\begin{itemize}
\item Step 1. Weight each edge by the two-particle interaction energy between the two vertices.
\item Step 2. Choose a random starting vertex.
\item Step 3. Generate an $N$-vertex walk with a minimum edge sum.
\end{itemize}

The subgraph formed by this walk specifies the vertices (particles) of the child cluster. This child cluster will not have the minimum possible energy since we start at a random vertex and we are neglect all other interactions. However, this method produces a child cluster with favourable genetics from both parents and so has a very high chance of being fitter than at least one parent.

\subsection{Mutation}
Another genetic operator called mutation \cite{genetic_operators} is applied, with a certain probability, to the children formed by crossover. This operator will take a child chromosome and randomly change a gene. In other words, we implement a function that takes in an individual and produces a new one that is slightly different with respect to the search space. Here, the individuals are elements of $\mathbb{R}^{3N}$ and so a possible mutation function is,
\begin{displaymath}
M(x) = x + \delta
\end{displaymath}
where $x,\delta \in \mathbb{R}^{3N}$. For sufficiently small $|\delta|$, $M(x)$ moves $x$ smoothly through $\mathbb{R}^{3N}$ in a smooth way such that the local neighbourhood of $x$ can be searched. The mutation function needs an element of randomness to ensure a diverse search range. This randomness is incorporated into the choice of $\delta$.

Let $\delta=(\delta_1,\delta_2,\delta_3,...,\delta_{3N})$ such that $\delta_i \sim \mathcal{N} (0,\sigma_i)$ for $1\leq N \leq 3N$. The standard deviation, $\sigma_i$, can be chosen in a general way for each $i$ and may also depend on population parameters such as: average fitness, generation number, etc. Adjusting the degree of mutation, and/or the crossover method, turns a simple GA into an AGA. This adaptive behaviour allows for a much broader region of interest in the beginning of the search and then focuses on a particular region of the search space closer to the end of the search.
\\

In summary, a genetic algorithm works by applying genetic operators (crossover and mutation) to the generations but changing the parameters of these functions or even the implementations over time could be beneficial. In the early stages of searching, it would be best to have a large and diverse population to adequately search a large portion of the search space and so large mutations would be needed but in the later stages of the search it would be better to "fine tune" the population to the optimal solution and so smaller mutations or even a different form of mutation may be needed. This thinking leads to the idea of an Adaptive Genetic Algorithm (AGA) \cite{AGA} which has the ability to change its parameters (selection bias, mutation rate, generation size, etc) or even the functions which implement crossover and mutation based on the current state of the population. AGAs can therefore extend or narrow the range of the search if needed and in doing so reach an optimal rate of convergence to the optimal solution. 

\section{Results and Analysis}
\label{sec3}
\begin{figure*}[hbt!]
	\centering
    \includegraphics[width=10cm]{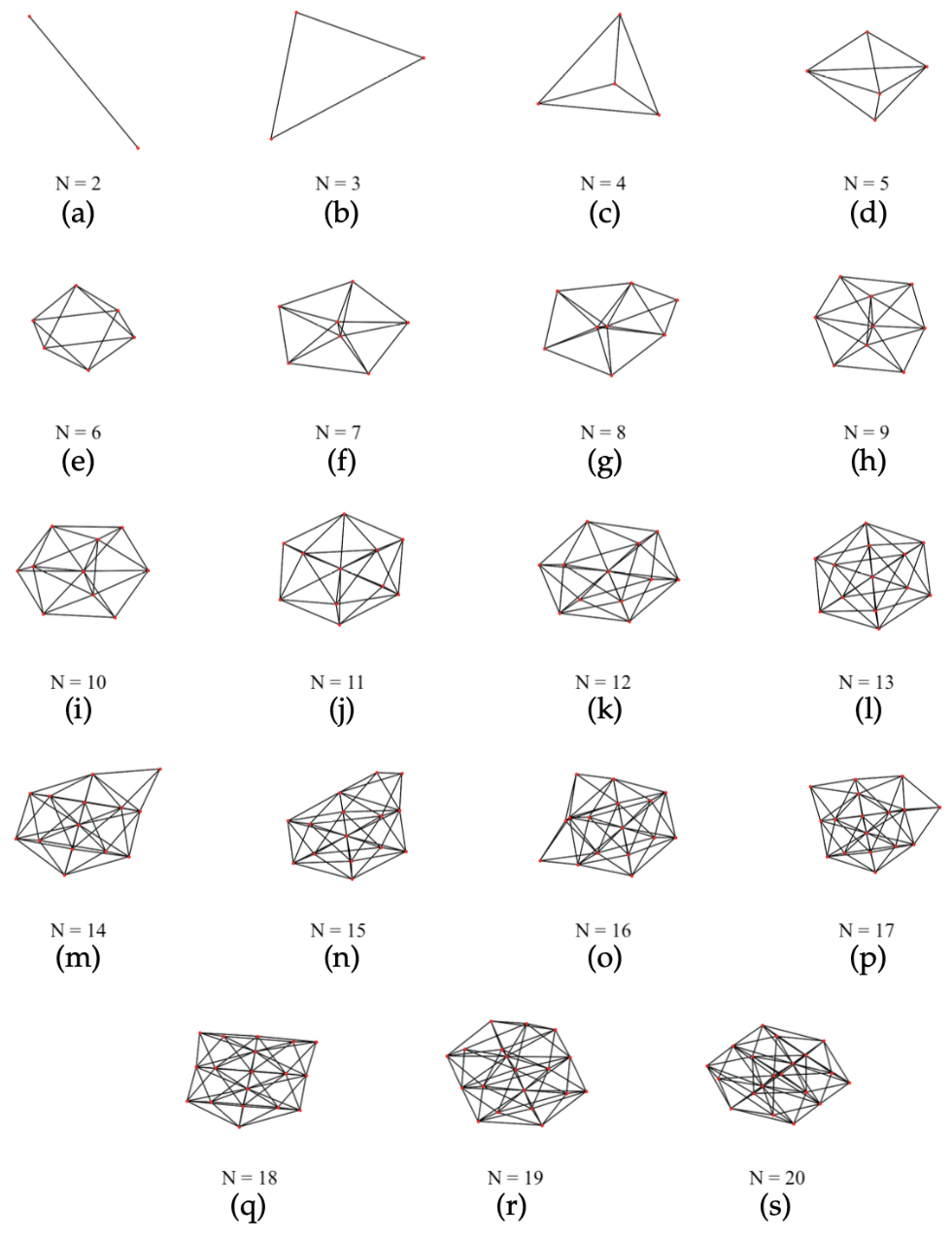}
\caption{The Lennard-Jones particle clusters for $N=2,3,...,20$}
\end{figure*}
The graphs of the particle clusters for $N=2,3,..,20$ are shown in Figure 2. The associated energy for each of the clusters is listed in Table 1. The values obtained from the AGA are compared to those values obtained from Ref. \cite{basin_hopping} where the basin hopping algorithm was used to find the global minimum cluster and associated energy. It should be noted that the values for selective cluster sizes match exactly, to the last decimal, with those found in Ref.\cite{Barron} and \cite{Muller} using a genetic algorithm approach.

\begin{table}[h]
	\caption{Table showing the minimum energy values found by the AGA and from Ref. [16].}
{\begin{tabular}{@{}ccc@{}} \toprule
		Particle Number ($N$) & Energy (This work) & Energy (\cite{basin_hopping}) \\ \colrule
		
		2	& -1.000000	 & -1.000000\\
		
		3	& -3.000000	 & -3.000000\\
		
		4	& -6.000000  & -6.000000\\
		
		5	& -9.103852415681363	 & -9.103852\\
		
		6	& -12.712062256782637 & -12.712062\\
		
		7	& -16.505384167507653 & -16.505384\\
		
		8	& -19.821489187804726 & -19.821489\\
		
		9	& -24.113360433066944 & -24.113360\\
		
		10	& -28.422531893437572 & -28.422532\\
		
		11	& -32.765970084618380 & -32.765970\\
		
		12	& -37.96759955862121 & -37.967600\\
		
		13	& -44.32680141873467 & -44.326801\\
		
		14	& -47.845156776950766 & -47.845157\\
		
		15	& -52.322627246105675 & -52.322627\\
		
		16	& -56.81574176701636 & -56.815742\\
		
		17	& -61.31799464848197 & -61.317995\\
		
		18	& -66.53094945127545 & -66.530949\\
		
		19	& -72.65978243384927 & -72.659782\\
		
		20	& -77.17704248805858 & -77.177043\\ \botrule
	\end{tabular} \label{tab:results}}
\end{table}
	
From Table 1, we see that the energy values obtained from our AGA match up to 6 decimals to that from Ref. \cite{basin_hopping} but our accuracy is very much superior since our precision is only limited to machine precision.
The performance of this AGA can be studied by looking at the rate of convergence for each of the clusters. This was recorded for particle numbers 5, 10, 15 and 20 as a sample set. It must be noted that this is not an objective measure due to the random nature of the AGA, however, it gives a good indication of how the AGA will perform and scale to larger numbers of particles. The rates of convergence are shown in Figure 3. The performance of these test runs are summarized in the Table 2.
\begin{figure*}[hbt!]
\centering
\includegraphics[width=12cm]{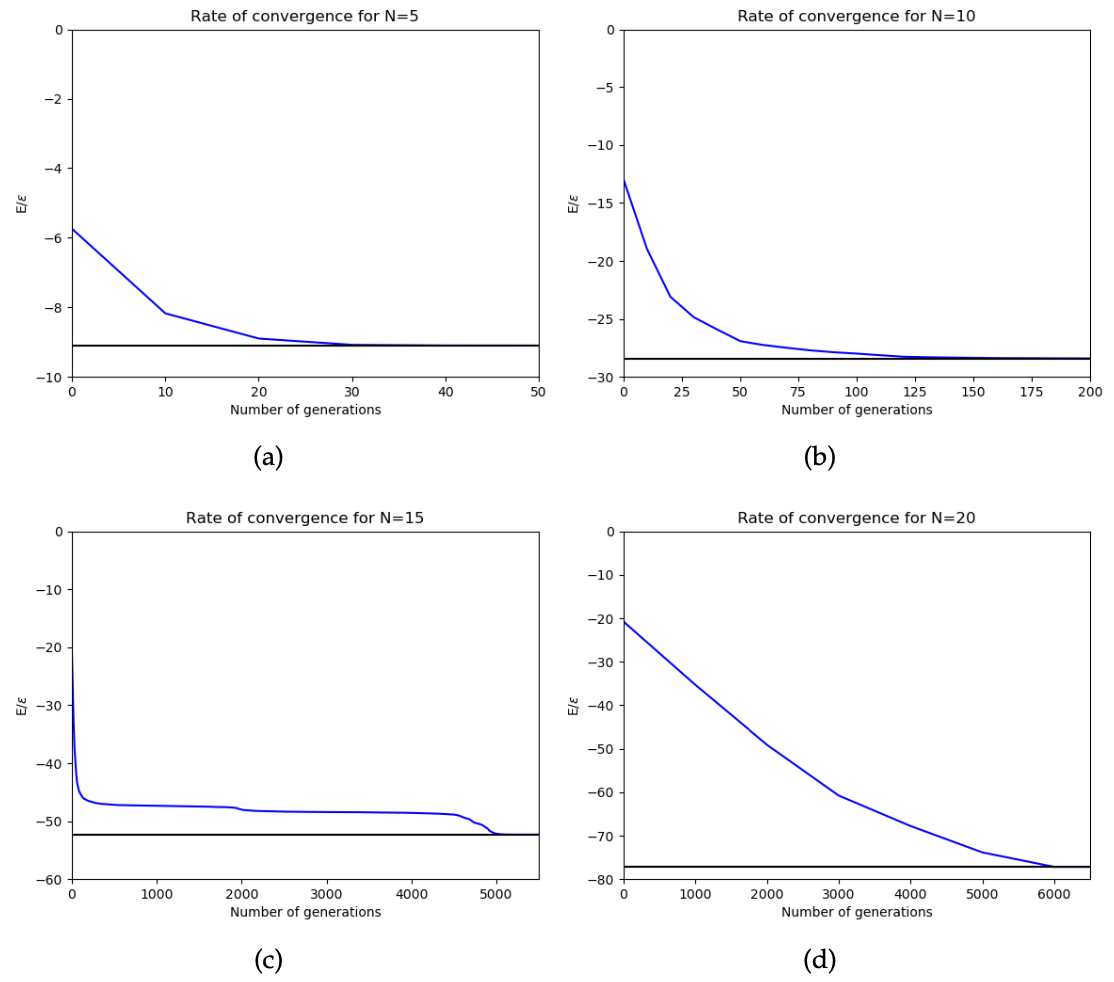}
	\caption{Plots of the AGA convergence for 5, 10, 15 and 20 particle clusters. The horizontal line shows the minimum energy for that particular cluster.}
\end{figure*}
\newline

 \begin{table}[h!]
	\caption{Table showing a summary of the performance of the AGA for the above mentioned sample set.}
{\begin{tabular}{@{}ccc@{}} \toprule
 		Particle Number & Number of generations & Convergence Point\\ \colrule
 		
 		5 & 50 & 40 \\
 		
 		10 & 200 & 150 \\
 		
 		15 & 5500 & 5000 \\
 		
 		20 & 6500 & 6000\\ \botrule
 		
 	\end{tabular}
 	\label{tab:performance}}
 \end{table} 
The sudden increase in the convergence point from 15 particles can be attributed to the intricate nature of the lowest energy cluster. The initialization of the generation is random apart from the fact that we enforce a minimum separation. This random start will then require many more generations to move through the configurations and avoid the many local minima. The number of generations can be reduced by specifying starting clusters that more resemble the lowest energy cluster, however, this can be very difficult for larger systems. 

\section{Conclusion}
\label{sec4}
A novel adaptive genetic algorithm was proposed and discussed in much detail. This AGA was designed to work for any pair potential and as an illustration, it was used to determine the lowest energy structures of various Lennard-Jones clusters with particle numbers in the range $2\leq N \leq 20$. The results surpasses the accuracy obtained from more elaborate methods including other methods employing genetic algorithms. In fact the precision of these results is only governed by the machine precision. The performance of the AGA was studied for a small sample set and some remarks were made. Work is currently in progress to improve the rate of convergence for this AGA as well as the scaling to larger number of particles.

\section*{Acknowledgments}
The author's acknowledge financial support from the National Research Foundation (NRF) of South Africa.

\end{document}